\documentclass[11pt,a4paper]{article}
\usepackage{amsmath,amsfonts,amssymb,amsthm}
\usepackage{fullpage,feynmp}

\newtheorem{thm}{Theorem}

\newtheorem{defn}[thm]{Definition}
\newtheorem{ex}[thm]{Example}
\newtheorem{rem}[thm]{Remark}
\def\C{\mathbb{C}} 
\def\gf{\mathrm{gf}}

\def\ii{\mathrm{i}}
\def\inter{\mathrm{int}}
\def\L{\mathcal{L}}
\def\N{\mathbb{N}} 
\def\P{\mathbb{P}}

\def\cO{\mathcal{O}}
\def\cQ{\mathcal{Q}}
\def\bar{\overline}
\def\tr{\mathrm{tr~}}
\def\tilde{\widetilde}
\newcommand{\ext}[1]{{(#1)}}

\title{The Hopf algebra of Feynman graphs in QED}
\author{Walter van Suijlekom\\[7mm]
Max Planck Institute for Mathematics\\
Vivatsgasse 7, D-53111 Bonn, Germany\\
\texttt{waltervs@mpim-bonn.mpg.de}}
\date{May 22, 2006}

\begin{document}
\begin{fmffile}{graphs}
\fmfcmd{%
style_def doublehead expr p =
  draw (wiggly subpath (0,8length(p)/10) of p);
  draw_double subpath (8length(p)/10,length(p)) of p
  enddef;}
\fmfset{wiggly_len}{5pt} 
\fmfset{wiggly_slope}{70} 

\fmfset{dot_len}{1mm}

\maketitle

\begin{abstract}
We report on the Hopf algebraic description of renormalization theory of quantum electrodynamics. The Ward-Takahashi identities are implemented as linear relations on the (commutative) Hopf algebra of Feynman graphs of QED. Compatibility of these relations with the Hopf algebra structure is the mathematical formulation of the physical fact that WT-identities are compatible with renormalization. As a result, the counterterms and the renormalized Feynman amplitudes automatically satisfy the WT-identities, which leads in particular to the well-known identity $Z_1=Z_2$.
\end{abstract}

\section{Introduction}
The description of perturbative quantum field theory in terms of Hopf algebras \cite{Kre98,CK99,CK00} (cf. \cite{CM04c,CM04b,CM06}) has led to a better understanding of the combinatorial structure of renormalization. At the same time, it provided a beautiful interaction of quantum field theory with several parts of mathematics. 

Although the Hopf algebra was modelled on a scalar quantum field theory, it can be generalized to any quantum field theory. However, subtleties appear in the case of quantum gauge theories. Gauge theories are field theories that are invariant under a certain group, known as the gauge group. In the process of renormalization, one has to deal with the `overcounting' due to the gauge degrees of freedom. Furthermore, one has to understand how the gauge symmetries manifest themselves in the quantum field theory. Important here are the so-called Ward-Takahashi (or Slavnov-Taylor) identities.

The first example of a gauge theory is quantum electrodynamics (QED), which has the abelian group $U(1)$ as a gauge group. The Hopf algebraic structure of Feynman graphs in QED was desrcibed in \cite{BK99,KD99} and more recently in \cite{VP04}. A slightly different approach is taken in \cite{BF01,BF03}, where Hopf algebras on planar binary trees were considered. 
The anatomy of a gauge theory with gauge group $SU(3)$ (known as quantum chromodynamics) has been discussed in the Hopf algebra setting in \cite{Kre05}, with a central role played by the Dyson-Schwinger equations (see also \cite{KY06}). However, a full understanding of renormalization of general gauge theories is still incomplete. This paper is a modest attempt towards understanding this by working out explicitly the Hopf algebraic structure underlying renormalization of quantum electrodynamics, thereby implementing the Ward-Takahashi identities in a compatible way.

\bigskip

We start in Section \ref{sect:hopf} by defining the commutative Hopf algebra $H$ of Feynman graphs for quantum electrodynamics. The approach we take differs from the one in \cite{BF03} where a noncommutative Hopf algebra was considered. This was motivated by the fact that in QED, Feynman amplitudes are (noncommuting) matrices. Instead, we work with a commutative Hopf algebra and encode the matricial form of the Feynman amplitudes into the distributions giving the external structure of the graph. Our approach has the advantage of still being able to use the duality between commutative Hopf algebras and affine group schemes in the formulation of the BPHZ-procedure.

In Section \ref{sect:birkhoff}, we associate a Feynman amplitude to a Feynman graph, as dictated by the Feynman rules. The BPHZ-formula of renormalization is obtained as a special case of the Birkhoff decomposition for affine group schemes \cite{CK99}. 

Then in Section \ref{sect:wt}, we incorporate the Ward-Takahashi (WT) identities as relations between graphs, much as was done by 't Hooft and Veltman in \cite{HV73}. We show that these relations are compatible with the coproduct in the Hopf algebra $H$, so that they define a Hopf ideal. This reflects the physical fact that WT-identities are compatible with renormalization (in the dimensional regularization and minimal subtraction scheme). 

The Feynman amplitudes can now be defined on the quotient Hopf algebra of $H$ by this ideal, and we conclude that the counterterms as well as the renormalized Feynman amplitudes satisfy the WT-identities in the physical sense ({i.e.}, between Feynman amplitudes). In particular, we arrive at the well-known identity $Z_1=Z_2$ as derived by Ward in \cite{War50}.

\section{Hopf algebra structure of Feynman graphs in QED}
\label{sect:hopf}
Quantum electrodynamics in 4 dimensions is given by the following classical Lagrangian,
\begin{align*}
\L=- \frac{1}{4} F_{\mu\nu} F^{\mu\nu} - \frac{1}{2\xi}(\partial^\mu A_\mu)^2 +\bar\psi \left( \gamma^\mu(\partial_\mu+ e A_\mu)-m\right) \psi,
\end{align*}
where $F_{\mu\nu}=\partial_\mu A_\nu - \partial_\nu A_\mu$ is the {\it field strength} of the {\it gauge potential} $A_\mu$, $\psi$ is the spinor describing the electron with mass $m$ and electric charge $e$ and $\gamma^\mu, \mu=1,\ldots,4$ are the $4 \times 4$ Dirac matrices. Finally, the real parameter $\xi$ is the {\it gauge-fixing parameter}. 

The Lagrangian $\L$ describes the dynamics and the interaction as well as the gauge fixing: we can write $\L=\L_0 + \L_\inter+\L_\gf$, where the free, interaction and gauge fixing parts are
\begin{align*}
\L_0&=A_\mu D_{\mu\nu} A_\nu + \bar\psi (\gamma^\mu \partial_\mu -m) \psi,\\
\L_\inter&= e \bar\psi A_\mu \psi.\\
\L_\gf &= -\frac{1}{2\xi}(\partial^\mu A_\mu)^2.
\end{align*}
The adjective ``quantum'' for electrodynamics is justified when this Lagrangian is used in computing probability amplitudes. In perturbation theory, one expands such amplitudes in terms of Feynman diagrams which are the graphs constructed from the following vertices, as dictated by the form of the Lagrangian:
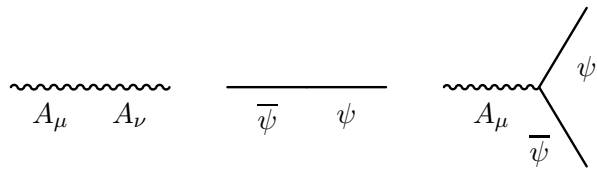
\begin{center}
    \begin{fmfgraph*}(60,60)
      \fmfleft{l}
      \fmfright{r}
      \fmf{photon,label=$A_\mu$}{l,v}
      \fmf{photon,label=$A_\nu$}{v,r}
    \end{fmfgraph*}\qquad
    \begin{fmfgraph*}(60,60)
      \fmfleft{l}
      \fmfright{r}
      \fmf{plain,label=$\bar\psi$}{l,v}
      \fmf{plain,label=$\psi$}{v,r}
    \end{fmfgraph*}\qquad
    \begin{fmfgraph*}(60,60)
      \fmfleft{l}
      \fmfright{r1,r2}
      \fmf{photon,label=$A_\mu$}{l,v}
      \fmf{plain,label=$\bar\psi$}{r1,v}
      \fmf{plain,label=$\psi$}{v,r2}
    \end{fmfgraph*}
\end{center}
\begin{rem}
\label{rem:el-mass}
In fact, we will also need the 2-point vertex 
\begin{fmfgraph*}(40,10)
      \fmfforce{(0w,.15h)}{l}
      \fmfforce{(1w,.15h)}{r}
      \fmf{plain}{l,v}
      \fmf{plain}{v,r}
      \fmfdot{v}
    \end{fmfgraph*}
, associated to the electron mass term, similar to \cite{BF01}. 
\end{rem}
We will focus on the so-called {\it one-particle irreducible} (1PI) {\it graphs}, which are graphs that are not trees, and that cannot be disconnected by cutting a single internal edge. In QED, there are three types of 1PI graphs that are of interest in renormalization theory: the vacuum polarization, the  electron self-energy and the full vertex graphs:
\begin{center}
    \begin{fmfgraph*}(60,60)
      \fmfleft{l}
      \fmfright{r}
      \fmflabel{$\Gamma(q,\mu,\nu)=q$}{l}
      \fmflabel{$q$}{r}
      \fmf{photon,label=$\mu$}{l,v}
      \fmf{photon,label=$\nu$}{v,r}
      \fmfblob{.25w}{v}
    \end{fmfgraph*}\\
    \begin{fmfgraph*}(60,60)
      \fmfleft{l}
      \fmfright{r}
      \fmflabel{$\Gamma(p)=p$}{l}
      \fmflabel{$p$}{r}
      \fmf{fermion}{l,v}
      \fmf{fermion}{v,r}
      \fmfblob{.25w}{v}
    \end{fmfgraph*}\\
    \begin{fmfgraph*}(60,60)
      \fmfleft{l}
      \fmfright{r1,r2}
      \fmflabel{$\Gamma(q,\mu;p)=q$}{l}
      \fmflabel{$p$}{r1}
      \fmflabel{$p+q$}{r2}
      \fmf{photon,label=$\mu$}{l,v}
      \fmf{fermion}{r1,v}
      \fmf{fermion}{v,r2}
      \fmfblob{.25w}{v}
    \end{fmfgraph*}
\end{center}
\bigskip
Here the blob stands for any 1PI graph of the type dictated by the external lines.

\bigskip

We now describe the {\it external structure} of a Feynman graph $\Gamma$, assigning (among other data) momenta to the external legs. In physics, the Feynman amplitude of $\Gamma$ (dictated by the Feynman rules, see below) is evaluated with respect to this external structure (see below). Mathematically, the external structure is a distribution on the space of Feynman amplitudes, understood as test functions on a suitable space. 

More explicitly, in the case of QED the (Euclidean) Feynman amplitudes $U(\Gamma)$ are functions in $C^\infty(E_\Gamma) \otimes M_4(\C)$ where
\begin{align*}
E_\Gamma=\left\{ (q_1,\ldots,q_n,\mu_1,\ldots,\mu_n,p_{n+1}, \ldots, p_N) : \sum p_i+q_j=0,  \right\}
\end{align*}
with $\{1,\ldots,n\}$ the set of indices labelling the external photon lines of $\Gamma$ with spatial index $\{\mu_1,\ldots,\mu_n\}$ respectively, and $\{n+1,\ldots,N\}$ is the set of its external electron lines. The factor $M_4(\C)$ can be understood from the fact that in QED, Feynman amplitudes are matrix-valued functions on $E_\Gamma$. 

For QED, the external structure is thus given by an element in the space of distributions on $C^\infty(E_\Gamma) \otimes M_4(\C)$; we denote this space by $\left(C^\infty(E_\Gamma) \otimes M_4(\C)\right)'$. For example, for the above full vertex graph $\Gamma$, $E_\Gamma=\left\{ (q,\mu,p,p'): \sum q+ p+p'=0 \right\}$ and a typical distribution is given by $\sigma^{kl}=\delta_{q,\mu,p,-p-q} \otimes e^{kl}$ in terms of a Dirac mass and the (standard) dual basis $\{ e^{kl}\}$ of $M_4(\C)'$. Evaluation on the Feynman amplitude $U(\Gamma) \in C^\infty(E_\Gamma)\otimes M_4(\C)$ is then given by the pairing
\begin{align*}
\langle \sigma^{kl}, U(\Gamma) \rangle = U(\Gamma)(q,\mu;p,-p-q)_{kl}
\end{align*}
Note that $p'=-p-q$ is translated in the diagram by a reversal of the corresponding arrow with associated momentum $p+q$. 

\bigskip

We will be interested in the following special external structures for the different graphs introduced above. A full vertex graph $\Gamma(q,\mu;p)$ at zero momentum transfer (meaning $q=0$) can be written in terms of the two form factors $F_1, F_2$ (cf. \cite[Section 6.2]{PS95})
\begin{align}
U(\Gamma)(q=0,\mu;p)=\gamma_\mu F_1(p^2) + \frac{\not{p} \gamma_\mu \not{p}}{p^2} F_2(p^2).
\end{align}
where $\not{p}=\sum_\mu \gamma^\mu p_\mu$. Only the first form factor $F_1$ requires renormalization; therefore, we define the following external structure:
\begin{align}
\label{ext1}
\langle \sigma_\ext{1},f \rangle &= \frac{1}{16}\sum_\mu \tr \gamma_\mu f(q=0,\mu;p=0),\qquad \forall f \in C^\infty(E_\Gamma) \otimes M_4(\C),
\end{align}
where the normalization comes from $\tr \gamma^\mu \gamma_\mu = 16$. For the Ward-Takahashi identities, we will also need the following external structure,
\begin{align}
\langle \sigma_\ext{0,\mu, p}^{kl}, f \rangle = f(q=0, \mu; p)_{kl},
\end{align}
which puts finite momentum $p$ on the ingoing and outgoing electron lines, but zero momentum on the external photon line (zero-momentum transfer). This external structure allows us to treat Ward-Takahashi identities in massless QED as well. 

\begin{rem}
The form factors $F_1$ and $F_2$ can also be recovered separately by using the following two distributions \cite{KD99},
\begin{align}
\langle \sigma',f \rangle &= \tr \sum_\mu \gamma^\mu f(q=0, \mu;p),  \\
\langle \sigma'',f \rangle &= \tr \sum_\mu \not{p} p_\mu f(q=0, \mu;p).
\end{align}
\end{rem}

\bigskip

For an electron self-energy graph $\Gamma(p)$, we have $E_\Gamma = \{ (p)\}$ and the corresponding Feynman amplitude is written in terms of two form factors:
\begin{align}
U(\Gamma)(p)=\not{p} A(p^2) + m B(p^2).
\end{align}
Correspondingly, there are two distributions, which we choose of the form,
\begin{align}
\label{ext0}
\langle \sigma_\ext{0},f \rangle &= \frac{1}{16} \sum_\mu \tr \gamma_\mu \frac{\partial}{\partial p_\mu} f(p) \big|_{p=0} +  m^{-1} ~\tr f(p=0),\\
\label{ext2}
\langle \sigma_\ext{2},f \rangle &= \frac{1}{16} \sum_\mu \tr \gamma_\mu \frac{\partial}{\partial p_\mu} f(p) \big|_{p=0},
\end{align}
for all $f \in C^\infty(E_\Gamma) \otimes M_4(\C)$. Also, there is an external structure which puts finite momentum on the ingoing and outgoing electron lines,
\begin{align}
\langle \sigma_\ext{p_\mu} , f \rangle = \frac{\partial}{\partial p_\mu} f(p)
\end{align}

Finally, for a vacuum polarization graph $\Gamma$, $E_\Gamma = \{ (q, q', \mu,\nu) : q+q'=0 \} \equiv \{ (q, \mu,\nu )\}$, we let $\sigma_\ext{3}$ be the external structure that satisfies
\begin{align}
\label{ext3}
\langle \sigma_\ext{3}, \left(- \delta_{\mu\nu}q^2+ q_\mu q_\nu\right)\otimes M \rangle = \tr M,
\end{align}
for all $M \in M_4(\C)$. 

\begin{rem}
The labelling of the external structures by $\ext{0}, \ext{1}, \ext{2}$ and $\ext{3}$ differs from the one in \cite{CK00}, where they were labelled by $\ext{0}$ and $\ext{1}$. Here we follow \cite{BF01}, so that the labels correspond to the renormalization constants $m_0, Z_1, Z_2, Z_3$, as they appear in the Lagrangian,
\begin{align}
\label{eq:L-ren}
\L =  - \frac{1}{4} Z_3 A_\mu D_{\mu\nu} A_\nu - \frac{1}{2\xi}(\partial^\mu A_\mu)^2 + Z_2 \bar\psi \left( \gamma^\mu \partial_\mu - m_0\right) \psi  + e Z_1 \bar\psi \gamma^\mu A_\mu \psi .
\end{align}
The precise correspondence will be given in equation \eqref{eq:ren-const} below.
\end{rem}

\bigskip

Let $H$ be the free commutative algebra generated by pairs $(\Gamma, \sigma)$ with $\Gamma$ a 1PI graph and $\sigma \in \left(C^{\infty} (E_\Gamma) \otimes M_4(\C) \right)'$ a distribution encoding its external structure. The product in $H$ can be understood as the union of graphs, with induced external structure. More precisely, for two 1PI graphs $\Gamma_\sigma$ and $\Gamma'_{\sigma'}$, the product $\Gamma_\sigma \cdot \Gamma'_{\sigma'}$ is the graph $\Gamma \cup \Gamma'$ with external structure given by $\sigma \otimes \sigma' \in \left(C^\infty(E_\Gamma \times E_{\Gamma'}) \otimes M_4(\C)\right)'$. The algebra $H$ is a (positively) graded algebra $H=\oplus_{n\in \N} H^n$ , with the grading given by the loop number of the graph. 

In the following, we will also write $\Gamma_\sigma$ for the pair $(\Gamma,\sigma)$, and sometimes even $\Gamma$. We will shorten the notation for graphs equipped with the special external structures defined above by writing $\Gamma_\ext{k} = (\Gamma, \sigma_\ext{k})$.

We define a coproduct $\Delta :H\to H \otimes H$ as follows; if $\Gamma$ is a 1PI graph, then one sets
\begin{align}
\label{coproduct}
\Delta \Gamma = \Gamma \otimes 1 + 1 \otimes \Gamma + \sum_{\gamma \subset \Gamma} \gamma_{(k)} \otimes \Gamma/\gamma_{(k)}.
\end{align}
Here $\gamma$ is a proper subset of the graph $\tilde\Gamma$ formed by the internal edges of $\Gamma$ ({i.e.} $0 \subsetneq \gamma \subsetneq \tilde \Gamma$). The connected components $\gamma'$ of $\gamma$ are 1PI graphs with the property that the set of edges of $\Gamma$ that meet $\gamma'$ have two or three elements. The sum runs over all multi-indices $k$, one index for each 1PI connected component of $\gamma$, and one denotes by $\gamma'_{(k)}$ the 1PI graph that has external structure as defined in equation \eqref{ext1}-\eqref{ext3}, with $k=3$ in the case of a vacuum polarization, $k=1$ for a full vertex graph and $k=0$ or $2$ in the case of an electron self-energy. In equation \eqref{coproduct}, $\gamma_{(k)}$ denotes the disjoint union of all graphs $\gamma'_{(k)}$ associated to the connected components of $\gamma$ and the graph $\Gamma/\gamma_{(k)}$ is the graph $\Gamma$ (with the same external structure) with each $\gamma'$ reduced to a vertex of type $(k)$. More explicitly, if $\gamma'=$
\begin{fmfgraph*}(30,11)
      \fmfforce{(0w,.15h)}{l}
      \fmfforce{(1w,.15h)}{r}
      \fmf{plain}{l,v}
      \fmf{plain}{v,r}
      \fmfblob{.25w}{v}
\end{fmfgraph*}, then for $k=2$ one replaces this graph in $\Gamma$ by the line     
\begin{fmfgraph*}(20,11)
      \fmfforce{(0w,.15h)}{l}
      \fmfforce{(1w,.15h)}{r}
      \fmf{plain}{l,v,r}
      \fmfv{decor.shape=cross,decor.size=2thick,label.dist=4pt,label.angle=90,label=\tiny{$\ext{2}$}}{v}
\end{fmfgraph*}
$ \equiv$
      \begin{fmfgraph*}(20,11)
      \fmfforce{(0w,.15h)}{l}
      \fmfforce{(1w,.15h)}{r}
      \fmf{plain}{l,v}
      \fmf{plain}{v,r}
\end{fmfgraph*}
, and for $k=0$ by 
\begin{fmfgraph*}(20,11)
      \fmfforce{(0w,.15h)}{l}
      \fmfforce{(1w,.15h)}{r}
      \fmf{plain}{l,v,r}
      \fmfv{decor.shape=cross,decor.size=2thick,label.dist=4pt,label.angle=90,label=\tiny{$\ext{0}$}}{v}
\end{fmfgraph*}
$ \equiv $
\begin{fmfgraph*}(20,11)
      \fmfforce{(0w,.15h)}{l}
      \fmfforce{(1w,.15h)}{r}
      \fmf{plain}{l,v}
      \fmf{plain}{v,r}
      \fmfdot{v}
\end{fmfgraph*} (cf. Remark \ref{rem:el-mass}).
In the case of the vacuum polarization and the full vertex graph, one replaces $\gamma'$ by the corresponding vertex. 

By complete analogy with \cite{CK99}, we find that with this coproduct, $H$ becomes a connected graded Hopf algebra, {i.e.} $H=\oplus_{n \in \N} H^n$, $H^0=\C$ and
\begin{align*} 
\Delta(H^n) = \sum_{k=0}^n H^k \otimes H^{n-k}.
\end{align*}
Indeed, $H^0$ consists of complex multiples of the empty graph, which is the unit in $H$, so that $H^0=\C 1$. Moreover, from general results on graded Hopf algebras, we obtain the antipode inductively, 
\begin{align}
\label{antipode}
S(X)=-X - S(X')X'',
\end{align}
where $\Delta(X) = X\otimes 1 + 1\otimes X+ \sum X' \otimes X''$.

We give a few examples in order to clarify the coproduct. 
\begin{align*}
\Delta\bigg(
\parbox{40pt}{\begin{fmfgraph*}(40,11)
      \fmfleft{l}
      \fmfright{r}
      \fmf{plain}{l,v1,v2,v3,v4,r}
      \fmf{photon,left,tension=0}{v1,v4}
      \fmf{photon,left,tension=0}{v2,v3}
\end{fmfgraph*}}
\bigg) &= 
\parbox{40pt}{\begin{fmfgraph*}(40,11)
      \fmfleft{l}
      \fmfright{r}
      \fmf{plain}{l,v1,v2,v3,v4,r}
      \fmf{photon,left,tension=0}{v1,v4}
      \fmf{photon,left,tension=0}{v2,v3}
\end{fmfgraph*}}
 \otimes 1 + 1 \otimes 
\parbox{40pt}{\begin{fmfgraph*}(40,11)
      \fmfleft{l}
      \fmfright{r}
      \fmf{plain}{l,v1,v2,v3,v4,r}
      \fmf{photon,left,tension=0}{v1,v4}
      \fmf{photon,left,tension=0}{v2,v3}
\end{fmfgraph*}}
+
\parbox{40pt}{
\begin{fmfgraph*}(40,11)
      \fmfleft{l}
      \fmfright{r}
      \fmf{plain}{l,v2,v3,r}
      \fmf{photon,left,tension=0}{v2,v3}
\end{fmfgraph*}}
_\ext{2}
\otimes
\parbox{40pt}{
\begin{fmfgraph*}(40,11)
      \fmfleft{l}
      \fmfright{r}
      \fmf{plain}{l,v1,v4,r}
      \fmf{photon,left,tension=0}{v1,v4}
\end{fmfgraph*}}
+
\parbox{40pt}{
\begin{fmfgraph*}(40,11)
      \fmfleft{l}
      \fmfright{r}
      \fmf{plain}{l,v2,v3,r}
      \fmf{photon,left,tension=0}{v2,v3}
\end{fmfgraph*}}
_\ext{0}
\otimes
\parbox{40pt}{
\begin{fmfgraph*}(40,11)
      \fmfleft{l}
      \fmfright{r}
      \fmf{plain}{l,v1,v3,v4,r}
      \fmfdot{v3}
      \fmf{photon,left,tension=0}{v1,v4}
\end{fmfgraph*}}
\end{align*}
\begin{multline*}
\Delta\bigg(
\parbox{50pt}{\begin{fmfgraph*}(50,20)
      \fmfleft{l}
      \fmfright{r}
      \fmf{plain}{l,v1,v2,v3,v4,v5,v6,r}
      \fmf{photon,left,tension=0}{v1,v5}
      \fmf{photon,right,tension=0}{v2,v6}
      \fmf{photon,left,tension=0}{v3,v4}
\end{fmfgraph*}}
\bigg) = \parbox{50pt}{\begin{fmfgraph*}(50,20)
      \fmfleft{l}
      \fmfright{r}
      \fmf{plain}{l,v1,v2,v3,v4,v5,v6,r}
      \fmf{photon,left,tension=0}{v1,v5}
      \fmf{photon,right,tension=0}{v2,v6}
      \fmf{photon,left,tension=0}{v3,v4}
\end{fmfgraph*}} \otimes 1
+ 
1 \otimes \parbox{50pt}{\begin{fmfgraph*}(50,20)
      \fmfleft{l}
      \fmfright{r}
      \fmf{plain}{l,v1,v2,v3,v4,v5,v6,r}
      \fmf{photon,left,tension=0}{v1,v5}
      \fmf{photon,right,tension=0}{v2,v6}
      \fmf{photon,left,tension=0}{v3,v4}
\end{fmfgraph*}}
+
\parbox{40pt}{
\begin{fmfgraph*}(40,11)
      \fmfleft{l}
      \fmfright{r}
      \fmf{plain}{l,v3,v4,r}
      \fmf{photon,left,tension=0}{v3,v4}
\end{fmfgraph*}}
_\ext{2}
\otimes 
\parbox{50pt}{
\begin{fmfgraph*}(50,20)
      \fmfleft{l}
      \fmfright{r}
      \fmf{plain}{l,v1,v2,v5,v6,r}
      \fmf{photon,left,tension=0}{v1,v5}
      \fmf{photon,right,tension=0}{v2,v6}
\end{fmfgraph*}}
\\[5mm]
+
\parbox{40pt}{
\begin{fmfgraph*}(40,11)
      \fmfleft{l}
      \fmfright{r}
      \fmf{plain}{l,v3,v4,r}
      \fmf{photon,left,tension=0}{v3,v4}
\end{fmfgraph*}}
_\ext{0}
\otimes 
\parbox{50pt}{
\begin{fmfgraph*}(50,20)
      \fmfleft{l}
      \fmfright{r}
      \fmf{plain}{l,v1,v2,v3,v5,v6,r}
      \fmf{photon,left,tension=0}{v1,v5}
      \fmf{photon,right,tension=0}{v2,v6}
      \fmfdot{v3}
\end{fmfgraph*}}
+ 
\parbox{50pt}{\begin{fmfgraph*}(50,20)
    \fmfforce{(.33w,0h)}{b}
    \fmfleft{l}
    \fmfright{r}
    \fmf{plain}{l,v1,v3,v5,v6,v7,r}
    \fmffreeze
    \fmf{photon}{b,v3}
    \fmf{photon,left,tension=0}{v5,v6}
    \fmf{photon,left,tension=0}{v1,v7}
    \end{fmfgraph*}}
_\ext{1}
\otimes 
\parbox{40pt}{
\begin{fmfgraph*}(40,11)
      \fmfleft{l}
      \fmfright{r}
      \fmf{plain}{l,v3,v4,r}
      \fmf{photon,left,tension=0}{v3,v4}
\end{fmfgraph*}}
+
\parbox{50pt}{\begin{fmfgraph*}(50,20)
    \fmfforce{(.66w,0h)}{b}
    \fmfleft{l}
    \fmfright{r}
    \fmf{plain}{l,v1,v2,v3,v6,v7,r}
    \fmffreeze
    \fmf{photon}{b,v6}
    \fmf{photon,left,tension=0}{v2,v3}
    \fmf{photon,left,tension=0}{v1,v7}
    \end{fmfgraph*}}
_\ext{1}
\otimes 
\parbox{40pt}{
\begin{fmfgraph*}(40,11)
      \fmfleft{l}
      \fmfright{r}
      \fmf{plain}{l,v3,v4,r}
      \fmf{photon,left,tension=0}{v3,v4}
\end{fmfgraph*}}
\end{multline*}

\section{Birkhoff decomposition and renormalization}
\label{sect:birkhoff}
In \cite{CK99}, Connes and Kreimer understood renormalization of perturbative quantum field theory in terms of a Birkhoff decomposition. In particular, using the dimensional regularization (dim-reg) and minimal subtraction scheme, they have proved that the BPHZ-formula is a special case of the Birkhoff decomposition of a loop on a small circle in the complex plane, centered at the dimension of the theory, and taking values in a pro-unipotent Lie group. Before applying this to the case of QED, we briefly recall the Birkhoff decomposition, while referring to \cite{CK99} and \cite{CM04b} for more details.

The Birkhoff decomposition provides a procedure to extract a finite value from a singular expression. More precisely, let $C \in \P^1(\C)$ be a smooth simple curve and let $C_\pm$ denote its two complements with $\infty \in C_-$. The Birkhoff decomposition of a loop $\gamma :C \to G$, taking values in a complex Lie group $G$, is a factorization of the form
\begin{align*}
\gamma(z) = \gamma_-(z)^{-1} \gamma_+(z) ,\qquad z \in \C
\end{align*}
where $\gamma_\pm$ are boundary values of holomorphic maps (denoted by the same symbol)
\begin{align*}
\gamma_\pm : C_\pm \to G.
\end{align*}
The normalization $\gamma_-(\infty)=1$ ensures uniqueness of the decomposition (if it exists). 

The evaluation $\gamma \to \gamma_+(z_0)\in G$ is a natural principle to extract a finite value from the (possibly) singular expression $\gamma(z_0)$. This gives a multiplicative removal of the pole part for a meromorphic loop $\gamma$, when we let $C$ be an infinitesimal circle centered at $z_0$. 

Existence of the Birkhoff decomposition has been established for prounipotent complex Lie groups in \cite{CK99}, and, more generally, in \cite{CM04b} in the setting of affine group schemes. Recall that affine group schemes are dual to commutative Hopf algebras in the following sense. For a commutative Hopf algebra $H$, we understand an affine group scheme as a functor $G:A \to G(A)$ from the category of commutative algebras to the category of groups in the following way. For a given algebra $A$, one defines the group $G(A)$ to be the set of all homomorphisms from $H$ to $A$, with product, inverse and unit given as the duals with respect to the coproduct, antipode and counit $\epsilon$ respectively, {i.e.}
\begin{align*}
\phi_1 * \phi_2(X) &= \langle \phi_1 \otimes \phi_2, \Delta(X) \rangle,\\
\phi^{-1}(X)&=\phi(S(X)),\\
e(X)&=\epsilon (X).
\end{align*}
The following result is the algebraic translation of the Birkhoff decomposition to (pro-unipotent) affine group schemes, expressed in terms of the commutative (connected graded) Hopf algebra $H$ underlying this affine group scheme. Its proof can be found in \cite{CK99} (see also \cite{CM04b}). We assume that $C$ is an infinitesimal circle centered at $0 \in \C$ and denote by $K$ the field of convergent Laurent series, with arbitrary radius of convergence and by $\cO$ the ring of convergent power series. Furthermore, we set $\cQ=z^{-1} \C([z^{-1}])$. 
\begin{thm}[Connes-Kreimer]
\label{birkhoff}
Let $\phi:H \to K$ be an algebra homomorphism from a commutative connected graded Hopf algebra $H$ to the field $K$ defined above. The Birkhoff decomposition of the corresponding loop is obtained recursively from the equalities,
\begin{align*}
\phi_-(X)&=-T \left( \phi(X)+\sum \phi_-(X') \phi(X'') \right),
\end{align*}
where we have written $\Delta(X) = X \otimes 1 + 1 \otimes X + \sum X' \otimes X''$ for $X \in H$ and $T$ is the projection on the pole part in $z$, and $\phi_+= \phi_- * \phi$, or explicitly, 
\begin{align*}
\phi_+(X)&=\phi(X) + \phi_-(X) + \sum \phi_-(X') \phi(X'').
\end{align*}
The maps $\phi_-$ and $\phi_+$ are homomorphisms from $H$ to $\cQ$ and $\cO$ respectively.
\end{thm}
The loop $\gamma : C \to G$ defined in terms of $\phi: H \to K$ by $\gamma(z)(X):=\phi(X)(z)$ therefore factorizes as $\gamma=\gamma_-^{-1} \gamma_+$ where $\gamma_\pm:C_\pm \to G$ are defined in like manner in terms of $\phi_\pm$. The fact that they are holomorphic maps on $C_+$ and $C_-$, respectively, follows from the properties that $\phi_+$ maps to $\cO$ whereas $\phi_-$ maps to $\cQ$. 

\bigskip

Let us see how this applies in the case of quantum electrodynamics and in particular, how it gives the BPHZ procedure of renormalization of it. Let us start by giving the Feynman rules for the graphs in our Hopf algebra, allowing us to associate a Feynman amplitude to a graph $\Gamma$. We will use dim-reg (see \cite[Ch.4]{Col84}) and obtain the regularized (bare) Feynman amplitudes as integrals
\begin{align}
\label{eq:feynm-ampl-pre}
U_\Gamma (q_1,\ldots , q_n,\mu_1,\ldots,\mu_n,p_{n+1},\ldots,p_N)(z)
=\int d^{4-z}k_1 \cdots d^{4-z} k_L ~I_\Gamma(q,\mu,p,k_1,\ldots,k_L)
\end{align}
If we work in the Euclidean setting, the integrand is given by the following Feynman rules:
\begin{enumerate}
\item Assign a factor $\left( -\frac{\delta_{\mu\nu}}{p^2+\ii \epsilon} + \frac{p_\mu p_\nu}{(p^2+\ii\epsilon)^2} (1-\xi) \right)$ to each internal photon line. 
\item Assign a factor $\frac{1}{\gamma^\mu p_\mu + m}$ to each internal electron line.
\item Assign a factor $e \gamma^\mu$ to each 3-point vertex. The apparent dependence on the index $\mu$ is resolved by summing over $\mu$ in combination with the attached photon line (having also an index $\mu$). 
\item Assign a factor $m$ to the 2-point vertex. 
\item Assign a momentum conservation rule to each vertex.
\end{enumerate}
In rules {\it 1.} and {\it 2.}, we have introduced the IR-regulators $\ii \epsilon$ in order to resolve divergences at small momenta.
\begin{ex}
Consider the following electron self-energy graph\\
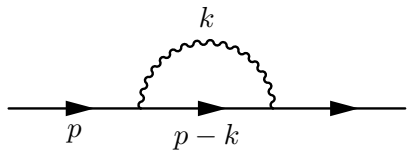
\begin{center}
\begin{fmfgraph*}(150,60)
      \fmfleft{l}
      \fmfright{r}
      \fmf{fermion,label=$p$}{l,v2}
      \fmf{fermion,label=$p-k$}{v2,v3}
      \fmf{fermion}{v3,r}
      \fmf{photon,left,tension=0,label=$k$}{v2,v3}
\end{fmfgraph*}
\end{center}
According to the Feynman rules, the integrand for this graph is 
\begin{align*}
I_\Gamma(p,k)=(e\gamma^\mu) \frac{1}{\gamma^\kappa (p_\kappa + k_\kappa) + m} (e\gamma^\nu) \left( -\frac{\delta_{\mu\nu}}{k^2+\ii \epsilon} + \frac{k_\mu k_\nu}{(k^2+\ii \epsilon)^2} (1-\xi) \right) 
\end{align*}
with summation over repeated indices understood. 
\end{ex}
As can be seen from equation \eqref{eq:feynm-ampl-pre}, a Feynman amplitude $U_\Gamma$ corresponding to a 1PI graph $\Gamma$ is an element in $C^\infty(E_\Gamma) \otimes M_4(\C) \otimes K$, and a map $U:H \to K$ can be defined by setting
\begin{align} 
\label{eq:feynm-ampl}
U (\Gamma_\sigma)(z)=e^{2-N} \langle \sigma, U_\Gamma \rangle.
\end{align} 
The factor $e^{2-N}$ is introduced in order to keep track of the loop number of the graph in terms of powers of $e^2$. Indeed, with $N$ the number of external edges of $\Gamma$, the power of $e^2$ appearing in the above expression is exactly the loop number $L$ of $\Gamma$. 

The counterterm $C(\Gamma)$ for a graph $\Gamma$ is given as the negative part of the Birkhoff decomposition of $U$ applied to $\Gamma$ and the renormalized value $R(\Gamma)$ as the positive part. Indeed, in this case, the above formul{\ae} give precisely the recursive BPHZ-procedure of subtracting divergences, dealing with possible subdivergences recursively, so that $C=U_-$ and $R=U_+$. More explicitly,
\begin{align*}
C(\Gamma) &= -T \left( 
U(\Gamma) + \sum_{\gamma \subset \Gamma} C(\gamma_{(k)}) U(\Gamma/\gamma_{(k)}) \right),\\
R(\Gamma) &=  
U(\Gamma) + C(\Gamma) + \sum_{\gamma \subset \Gamma} C(\gamma_{(k)}) U(\Gamma/\gamma_{(k)}).
\end{align*}

\begin{rem}
The relation between the renormalization constants in the Lagrangian (see equation \eqref{eq:L-ren}) and the counterterms, was given by Dyson in \cite{Dys49}. In terms of an expansion of Feynman graphs of the specified type, they read (compare with equation (73)-(76) in \cite{BF01})
\begin{equation}
\begin{aligned}
\label{eq:ren-const}
m_0 &= Z_2 m + \sum_{\Gamma=~
\parbox{20pt}{
\begin{fmfgraph}(20,10)
      \fmfleft{l}
      \fmfright{r}
      \fmf{plain}{l,v}
      \fmf{plain}{v,r}
      \fmfblob{.25w}{v}
\end{fmfgraph}}} 
    C(\Gamma_\ext{0}),\\
Z_1 &= 1 + \sum_{\Gamma=~
\parbox{20pt}{
\begin{fmfgraph}(20,10)
      \fmfleft{l}
      \fmfright{r1,r2}
      \fmf{photon}{l,v}
      \fmf{plain}{r1,v}
      \fmf{plain}{v,r2}
      \fmfblob{.25w}{v}
\end{fmfgraph}}} C(\Gamma_\ext{1}),\\
Z_2 &= 1 - \sum_{\Gamma=~
\parbox{20pt}{
\begin{fmfgraph}(20,10)
      \fmfleft{l}
      \fmfright{r}
      \fmf{plain}{l,v}
      \fmf{plain}{v,r}
      \fmfblob{.25w}{v}
\end{fmfgraph}}} C(\Gamma_\ext{2}),\\
Z_3 &=1 - \sum_{\Gamma=~
\parbox{20pt}{
\begin{fmfgraph}(20,10)
      \fmfleft{l}
      \fmfright{r}
      \fmf{photon}{l,v}
      \fmf{photon}{v,r}
      \fmfblob{.25w}{v}
\end{fmfgraph}}} C(\Gamma_\ext{3}).
\end{aligned}
\end{equation}
\end{rem}

\section{Ward-Takahashi identities}
\label{sect:wt}
In quantum electrodynamics, the Ward-Takahashi identities give relations between Feynman amplitudes for full vertex graphs and electron self-energy graphs. Our goal is to translate these identities into relations on the Hopf algebra $H$, thus giving relations between Feynman graphs. More precisely, we consider a quotient of the Hopf algebra $H$ by a Hopf ideal generated by so-called {\it Ward-Takahashi (WT) elements}, thereby establishing the well-known fact (see for example \cite{Col84, PS95}) that the WT-identities are compatible with renormalization. 

First, we introduce the following notation. Let $\Gamma$ be an electron self-energy graph, and number the internal electron lines that are not part of an electron loop by $i$. We define $\Gamma(i)$ to be the graph $\Gamma$ with insertion of a photon line on the $i$'th electron line.

\begin{defn}
For every electron self-energy graph $\Gamma$, we define the {\rm Ward-Takahashi element} $W(\Gamma) \in H$ associated to $\Gamma$ by
\begin{align*}
W(\Gamma)(p,q)_{kl} := \sum_i \sum_\mu q_\mu \Gamma(i)(q,\mu;p)_{kl} + \Gamma(p+q)_{kl} - \Gamma(p)_{kl}.
\end{align*}
The first term is understood as the pair $(\Gamma(i),\sigma) \in H$, with external structure given by the distribution $\sigma^{kl}=\sum q^\mu \delta_{q,\mu,p,-p-q} \otimes e^{kl}$. 
Moreover, we set $W'_\mu(\Gamma)(p) := \frac{\partial}{\partial q_\mu} W(\Gamma)(p,q) \big|_{q=0}$ and $W''(\Gamma):=\frac{1}{16} \sum_\mu \tr \gamma_\mu \frac{\partial}{\partial q_\mu} W(\Gamma)(p,q) \big|_{p=q=0}$; in other words
\begin{align*}
W'_\mu(\Gamma)(p) &= \sum_i \Gamma(i)_\ext{0,\mu,p} + \Gamma_\ext{p_\mu},\\
W''(\Gamma) &= \sum_i \Gamma(i)_\ext{1} + \Gamma_\ext{2}.
\end{align*}
\end{defn}
\noindent In what follows, we will suppress the matrix indices $k,l$ in $W(\Gamma)$ for notational convenience.

The following notation due to 't Hooft and Veltman of double headed photon lines provides a compact way to denote the above external structure in terms of diagrams:
\begin{align*}
\parbox{40pt}{\begin{fmfgraph*}(40,40)
      \fmfleft{l}
      \fmfright{r1,r2}
      \fmflabel{$p$}{r1}
      \fmf{plain}{r1,v}
      \fmf{plain}{v,r2}
      \fmf{doublehead,label=$q$}{l,v}
      \fmfblob{.25w}{v}
\end{fmfgraph*}}
= \sum q^\mu ~\parbox{40pt}{
\begin{fmfgraph*}(40,40)
      \fmfleft{l}
      \fmfright{r1,r2}
      \fmflabel{$p$}{r1}
      \fmf{photon,label=$(\noexpand q,,\noexpand \mu)$}{l,v}
      \fmf{plain}{r1,v}
      \fmf{plain}{v,r2}
      \fmfblob{.25w}{v}
\end{fmfgraph*}}\\[3mm]
\end{align*}
In diagrams, this leads to the familiar expressions for the WT-identities (see for example \cite[Ch. 7]{PS95}),
\begin{align*}
W(\Gamma)(p,q)&= \sum_{\parbox{40pt}{\centering \tiny{insertion\\ points}}
} 
\parbox{60pt}{
\begin{fmfgraph*}(60,60)
      \fmfleft{l}
      \fmfright{r1,r2}
      \fmf{doublehead, label=$q$}{l,v}
      \fmf{fermion,label=$p$}{r1,v}
      \fmf{fermion}{v,r2}
      \fmfblob{.25w}{v}
\end{fmfgraph*}}
+
\parbox{60pt}{
\begin{fmfgraph*}(60,60)
      \fmfleft{l}
      \fmfright{r}
      \fmf{fermion,label=$p+q$}{l,v}
      \fmf{fermion}{v,r}
      \fmfblob{.25w}{v}
\end{fmfgraph*}}
-
\parbox{60pt}{
\begin{fmfgraph*}(60,60)
      \fmfleft{l}
      \fmfright{r}
      \fmf{fermion,label=$p$}{l,v}
      \fmf{fermion}{v,r}
      \fmfblob{.25w}{v}
\end{fmfgraph*}}
\\[5mm]
W_\mu'(\Gamma)(p)&= \sum_{\parbox{40pt}{\centering \tiny{insertion\\ points}}} 
\parbox{60pt}{
\begin{fmfgraph*}(60,60)
      \fmfleft{l}
      \fmfright{r1,r2}
      \fmflabel{$p$}{r1}
      \fmf{photon,label=$(\noexpand 0,,\noexpand \mu)$}{l,v}
      \fmf{fermion}{r1,v}
      \fmf{fermion}{v,r2}
      \fmfblob{.25w}{v}
\end{fmfgraph*}}
+\frac{\partial}{\partial p_\mu}~
\parbox{60pt}{
\begin{fmfgraph*}(60,60)
      \fmfleft{l}
      \fmfright{r}
      \fmf{fermion,label=$p$}{l,v}
      \fmf{fermion}{v,r}
      \fmfblob{.25w}{v}
\end{fmfgraph*}}
\\[5mm]
W''(\Gamma)&= \sum_{\parbox{40pt}{\centering \tiny{insertion\\ points}}} 
\parbox{40pt}{
\begin{fmfgraph*}(40,40)
      \fmfleft{l}
      \fmfright{r1,r2}
      \fmf{photon}{l,v}
      \fmf{plain}{r1,v}
      \fmf{plain}{v,r2}
      \fmfblob{.25w}{v}
\end{fmfgraph*}}_\ext{1}
+
\parbox{40pt}{
\begin{fmfgraph*}(40,30)
      \fmfleft{l}
      \fmfright{r}
      \fmf{plain}{l,v}
      \fmf{plain}{v,r}
      \fmfblob{.25w}{v}
\end{fmfgraph*}}_\ext{2}
\end{align*}
for $\Gamma=$ \begin{fmfgraph*}(40,11)
      \fmfforce{(0w,.25h)}{l}
      \fmfforce{(1w,.25h)}{r}
      \fmf{plain}{l,v}
      \fmf{plain}{v,r}
      \fmfblob{.25w}{v}
\end{fmfgraph*}
any electron self-energy graph. 

\begin{thm}
\label{thm:WT}
The ideal generated (in $H$) by $W(\Gamma)$, $W_\mu'(\Gamma)$ and $W''(\Gamma)$ for all 1PI electron self-energy graphs $\Gamma$ is a Hopf ideal.
\end{thm}
Before stating the proof of this, we remark that it is thus possible to define the quotient Hopf algebra $\tilde H$ of $H$, with the induced coproduct, counit and antipode.
\begin{proof}
Recall that an ideal $I$ in a Hopf algebra $H$ is called a Hopf ideal if
$$
\Delta(I) \subseteq I \otimes H + H\otimes I, \qquad \epsilon(I)=0, \qquad S(I) \subseteq I.
$$
In our case of a connected graded Hopf algebra $H$, the last property follows from the first. Indeed, since $S$ is given inductively by equation \eqref{antipode}, we find that $S(I)\subseteq I$.

We set $\Delta'(X) := \Delta(X) - X \otimes 1 + 1 \otimes X= \sum X' \otimes X''$ and denote the ideal generated by the WT-identities by $I$ . Clearly, it is enough to establish $\Delta'(I) \subseteq I \otimes H + H \otimes I$. Moreover, since $\Delta$ is an algebra map by definition, it is enough to establish $\Delta' (W(\Gamma)) \in I \otimes H + H \otimes I$ and the analogous expressions for $W'_\mu(\Gamma)$ and $W''(\Gamma)$ for any electron self-energy graph $\Gamma$. 

\bigskip

Let us start by illustrating the general argument in the following special case of the self-energy graph $\Gamma$ displayed in Figure \ref{figure-blocks}.
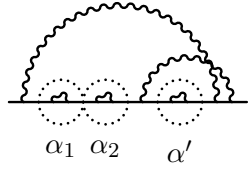
\begin{figure}[h!]
\begin{center}
\begin{fmfgraph*}(90,60)
      \fmfforce{(0w,.15h)}{l}
      \fmfforce{(1w,.15h)}{r}
      \fmf{plain}{l,v1,e12,v2,v3,e34,v4,v5,e56,v6,e67,v7,v8,e89,v9,v10,r}
      \fmffreeze
      \fmf{photon,left,tension=0}{v1,v9}
      \fmf{photon,left,tension=0}{v2,v3}
      \fmf{photon,left,tension=0}{v4,v5}
      \fmf{photon,left,tension=0}{v6,v10}
      \fmf{photon,left,tension=0}{v7,v8}
      \fmffreeze
      \fmf{dots,right,tension=0,label=$\alpha_1$}{e12,e34}
      \fmf{dots,right,tension=0}{e34,e12}
      \fmf{dots,right,tension=0,label=$\alpha_2$}{e34,e56}
      \fmf{dots,right,tension=0}{e56,e34}
      \fmf{dots,right,tension=0,label=$\alpha'$}{e67,e89}
      \fmf{dots,right,tension=0}{e89,e67}
\end{fmfgraph*}
\end{center}
\caption{An electron self-energy graph with blocks $\alpha=\alpha_1\cdot \alpha_2$ and $\alpha'$.}
\label{figure-blocks}
\end{figure}
We label the internal electron lines from left to right, from 1 to 9. 
Let us consider the first term in $W(\Gamma)$, and split its image under $\Delta'$ in a term in which $\gamma$ contains the electron line $i$ and a term in which it does not:
$$
\Delta'\left( \sum_i \Gamma(i) \right)= \sum_{\gamma \subset\Gamma} C(\gamma) + D(\gamma),
$$
with $C(\gamma):=  \sum_{i \in \gamma} \sum_k \gamma(i)_\ext{k} \otimes \Gamma(i)/\gamma(i)_\ext{k}$ and $D(\gamma):=\sum_{i\notin \gamma} \sum_k \gamma_\ext{k} \otimes \Gamma(i)/\gamma_\ext{k}$. 

For example, if $\gamma=\alpha_1 \cdot \alpha_2$ forms the `block' of concatenated electron self-energy graphs inside $\Gamma$, we find,
\begin{align*}
C(\gamma) &= \sum_{i=2,4} \sum_{k_1,k_2} \gamma(i)_\ext{k_1,k_2} \otimes \Gamma(i)/\gamma(i)_\ext{k_1,k_2}\\
&=\sum_k
(\tilde\alpha_1 \cdot \alpha_2)_\ext{1,k} \otimes \Gamma/\gamma_\ext{1,k}(e_1)
 + (\alpha_1\cdot\tilde\alpha_2)_\ext{k,1} \otimes \Gamma/\gamma_\ext{k,1}(e_2),
\end{align*}
where for $m=1,2$, $\tilde \alpha_m$ denotes the electron self-energy graph $\alpha_m$ with an external photon line attached to its internal electron line and $e_m$ is the electron line that corresponds to $\alpha_m$ in the quotient $\Gamma/\gamma$. On the other hand,
\begin{align*}
D(\gamma) &=\sum_{i=1,3,5,\ldots,9} \sum_{k_1,k_2} \gamma_\ext{k_1,k_2} \otimes \Gamma(i)/\gamma_\ext{k_1,k_2} \\
&= \sum_k (\alpha_1 \cdot \alpha_2)_\ext{2,k} \otimes \Gamma / \gamma_\ext{2,k}(1)
+ (\alpha_1 \cdot\alpha_2)_\ext{k,2} \otimes \Gamma / \gamma_\ext{k,2}(3) \\
&+\sum_k (\alpha_1 \cdot \alpha_2)_\ext{0,k} \otimes \Gamma / \gamma_\ext{0,k}(1)
+ (\alpha_1 \cdot \alpha_2)_\ext{k,0} \otimes \Gamma / \gamma_\ext{k,0}(3) \\
& +  \sum_{k_1,k_2} \gamma_\ext{k_1,k_2} \otimes \sum_{i=5,\ldots,9} \Gamma/\gamma_\ext{k_1,k_2}(i).
\end{align*}
Here we wrote explicitly the external structure $k=0,2$ for the electron self-energy graph in $\gamma$ for which an external photon line is attached to its incoming electron line (i.e. for $\alpha_1$ if $i=1$ and for $\alpha_2$ if $i=3$). Also, in the last line we used a labelling of the internal electron lines of the quotient $\Gamma/\gamma$ in terms of the electron lines of $\Gamma$ when $i \notin \gamma$.

The first two terms of $D(\gamma)$ combine with $C(\gamma)$ forming the WT-elements $W''(\alpha_m)$, and also the last three terms of $D(\gamma)$ combine to give,
\begin{align*}
C(\gamma)+D(\gamma)=&\sum_k W''(\alpha_1) (\alpha_2)_\ext{k} \otimes \Gamma/\gamma_\ext{2,k} (e_1)
+(\alpha_1)_\ext{k} W''(\alpha_2) \otimes \Gamma/\gamma_\ext{k,2} (e_2)\\
&+ \sum_{k_1,k_2} \gamma_\ext{k_1,k_2} \otimes \sum_{\parbox{45pt}{\centering \tiny{$j$ insertion \\points in $\Gamma/\gamma$}}} \Gamma/\gamma_\ext{k_1,k_2}(j),
\end{align*}
using for the last term the fact that the quotient $\Gamma/(\alpha_m)_\ext{0}$ gives rise to a 2-point vertex, and thus to a new internal electron line. It is easy to see that the first two terms in the above equation are in $I \otimes H$.

The two other terms in $\Delta'(W(\Gamma))$ are of the form $\sum_\gamma \gamma_{k_1,k_2} \otimes \Gamma/\gamma_\ext{k_1,k_2}$ and combine, for fixed $\gamma=\alpha_1 \cdot \alpha_2$ as above, with the last term in the previous equation to give precisely $\gamma_\ext{k_1,k_2} \otimes W(\Gamma/\gamma_\ext{k_1,k_2}) \in H \otimes I$. We conclude that the term in $\Delta'(W(\Gamma))$ arising from the subgraph $\gamma=\alpha_1\cdot \alpha_2$ is an element in $I \otimes H + H \otimes I$.
Similar arguments show that such relations hold for all subgraphs of $\Gamma$.

\bigskip

Let us now turn to the proof of the claim that $\Delta' (W(\Gamma)) \in I \otimes H + H \otimes I$ for any electron self-energy graph $\Gamma$. For the first term in $W(\Gamma)$ we have
\begin{align*}
\Delta'\left( \sum_i \Gamma(i) \right)=
\Delta'\left( \sum_i
\parbox{40pt}{\begin{fmfgraph*}(40,40)
      \fmfleft{l}
      \fmfright{r1,r2}
      \fmf{plain}{r1,v}
      \fmf{plain}{v,r2}
      \fmf{doublehead}{l,v}
      \fmfblob{.25w}{v}
\end{fmfgraph*}}
\right) = 
\sum_i \sum_{\gamma \subset \Gamma(i)}  \gamma_\ext{k} \otimes \Gamma(i)/\gamma_\ext{k},
\end{align*}
where $k$ is a multi-index labelling the external structure on the connected components of $\gamma$ and $i$ runs over all internal electron lines of $\Gamma$ that are not part of a loop. Also, we suppressed the external structure of $\Gamma(i)$ as it appears in $W(\Gamma)(p,q)$.

In general, $\gamma$ can be written as a union $\gamma=\gamma_E\cdot \gamma_P \cdot \gamma_V \cdot \gamma_E(i)$, where $\gamma_E$ consists of electron self-energy graphs, $\gamma_V$ of full vertex graphs, $\gamma_P$ of vacuum polarization graphs and $\gamma_E(i)$ of electron self-energy graphs with a photon line inserted at the $i$'th electron line (with this labelling inherited from $\Gamma$). Graphs of the type $\gamma_P(i)$ and $\gamma_V(i)$ do not appear in the sum, because there are no corresponding vertices of valence 3 and 4 in QED. 
Note that the multi-index $k$ only affects the subgraphs in $\gamma_E$, for each of which it can take the values $0$ and $2$. We split the above sum in a term in which $\gamma_\ext{k}$ contains the electron line $i$ as an internal line and those in which it does not:
\begin{align}
\label{cop-i:2}
\Delta'\left( \sum_i \Gamma(i) \right)= \sum_{\gamma \subset \Gamma}
\left(
\sum_{i\in \gamma_E} \gamma(i)_\ext{k} \otimes \Gamma(i)/\gamma(i)_\ext{k}
+\sum_{i\notin \gamma_E} \gamma_\ext{k} \otimes \Gamma(i)/\gamma_\ext{k}
\right),
\end{align}
where $\gamma=\gamma_E \cdot \gamma_V \cdot \gamma_P$. We will write $\Delta'\left( \sum_i \Gamma(i) \right)=\sum_\gamma C(\gamma) + D(\gamma)$ for the two terms appearing in equation \eqref{cop-i:2}.

Before examining the terms $C$ and $D$, we note that for each $\gamma$, $\gamma_E$ can be split into `blocks' each of which appear in the graph $\Gamma$ by concatenation of electron self-energy graphs (as in Figure \ref{figure-blocks}). We will denote such blocks by $\alpha \subset \gamma_E$ and write $C(\gamma)$ as
\begin{align*}
C(\gamma)= 
\sum_{\parbox{60pt}{\centering \tiny{$\alpha \subset \gamma_E$\\ $\alpha=\alpha_1 \cdots \alpha_l$}}} \sum_{m=1}^l \sum_{i \in \alpha_m} \sum_k \alpha_m(i)_\ext{1} ~(\gamma-\alpha_m)_\ext{k} \otimes 
\Gamma / \gamma_\ext{k} \left( e_m \right)
\end{align*}
where $e_m$ is the electron line corresponding to $\alpha_m$ in the quotient $\Gamma/\gamma$. In words, we separated the sum over the internal electron lines $i$ of $\gamma_E$ into a sum over all blocks $\alpha$ in $\gamma_E$, together with a sum over its connected components $\alpha_m$ and a sum over the internal electron lines that are part of $\alpha_m$. Then, with $i \in \alpha_m$, the quotient $\Gamma(i)/\gamma(i)_\ext{k}$ is given by the replacement of $\alpha_m$ by a 3-vertex, followed by the quotient by the other graphs that constitute $\gamma$, and with external structure inherited from $\Gamma(i)$.
Note that since $\alpha_m(i)$ is a full vertex graph, there is only the external structure $\alpha_m(i)_\ext{1}$. This fixes $k_m$ inside the multi-index $k$ to be $2$, and by a slight abuse of notation, we let $k$ also denote the external structures of the elements in $(\gamma-\alpha_m)$. 

The term $D(\gamma)$ can be split in a sum over $i$ of electron lines that are external edges for $\gamma_E$ and those that are not,
\begin{align*}
D(\gamma)= \sum_{i \in \partial \gamma_E} \sum_k \gamma_\ext{k} \otimes \Gamma(i)/\gamma_\ext{k} + \sum_{i \notin \gamma \cup \partial \gamma_E} \sum_k \gamma_\ext{k} \otimes \Gamma(i)/\gamma_\ext{k}.
\end{align*}
Again, we will write this in terms of the blocks $\alpha$ forming $\gamma_E$:
\begin{equation*}
\begin{aligned}
D(\gamma)&= \sum_\alpha \sum_{m=1}^l \sum_k 
(\alpha_m)_\ext{k_m} (\gamma-\alpha_m)_\ext{k} \otimes 
\Gamma
\left( \alpha_m \to 
  \parbox{20pt}{
    \begin{fmfgraph*}(20,20)
      \fmfleft{l}
      \fmfright{r}
      \fmfbottom{b}
      \fmf{plain}{l,v1,v2,v3,r}
      \fmffreeze
      \fmf{doublehead}{b,v2}
      \fmffreeze
      \fmfv{decor.shape=cross, decor.size=2thick,label.dist=4pt,label.angle=90,label=\tiny{$(k_m)$}}{v3}
    \end{fmfgraph*}}
\right)
/ 
 (\gamma-\alpha_m)_\ext{k}
\\
&+\sum_\alpha \sum_k 
(\alpha_l)_\ext{k_l} (\gamma-\alpha_l)_\ext{k} \otimes 
\Gamma
\left( \alpha_l \to 
  \parbox{20pt}{
    \begin{fmfgraph*}(20,20)
      \fmfleft{l}
      \fmfright{r}
      \fmfbottom{b}
      \fmf{plain}{l,v1,v2,v3,r}
      \fmffreeze
      \fmf{doublehead}{b,v2}
      \fmffreeze
      \fmfv{decor.shape=cross,decor.size=2thick,label.dist=4pt,label.angle=90,label=\tiny{$(k_l)$}}{v1}
    \end{fmfgraph*}}
\right)
/
 (\gamma-\alpha_l)_\ext{k} 
+ \sum_{i \notin \gamma \cup \partial \gamma_E}\sum_k \gamma_\ext{k} \otimes \Gamma(i)/\gamma_\ext{k},
\end{aligned}
\end{equation*}
where the first term arises from $i$ being the ingoing line of $\alpha_m$ for $m=1,\cdots l$ and the second term from $i$ being the outgoing line for $\alpha_l$, thus equal to the sum of all external edges in $\alpha$. 

\medskip

Writing explicitly the two terms for $k_m=0,2$ one obtains, 
\begin{equation*}
\begin{aligned}
D(\gamma)&= \sum_\alpha \bigg(\sum_{m=1}^l \sum_k 
(\alpha_m)_\ext{2} (\gamma-\alpha_m)_\ext{k} \otimes 
\Gamma/ \gamma_\ext{k} (e_m)
\\
&+
\sum_{m=1}^l \sum_k 
(\alpha_m)_\ext{0} (\gamma-\alpha_m)_\ext{k} \otimes 
\Gamma
\left( \alpha_m \to 
  \parbox{20pt}{
    \begin{fmfgraph*}(20,20)
      \fmfleft{l}
      \fmfright{r}
      \fmfbottom{b}
      \fmf{plain}{l,v1,v2,v3,r}
      \fmffreeze
      \fmf{doublehead}{b,v2}
      \fmffreeze
      \fmfv{decor.shape=cross, decor.size=2thick,label.dist=4pt,label.angle=90,label=\tiny{$\ext{0}$}}{v3}
    \end{fmfgraph*}}
\right)
/
 (\gamma-\alpha_m)_\ext{k} 
\\
&+\sum_k 
(\alpha_l)_\ext{k_l} (\gamma-\alpha_l)_\ext{k} \otimes 
\Gamma
\left( \alpha_l \to 
  \parbox{20pt}{
    \begin{fmfgraph*}(20,20)
      \fmfleft{l}
      \fmfright{r}
      \fmfbottom{b}
      \fmf{plain}{l,v1,v2,v3,r}
      \fmffreeze
      \fmf{doublehead}{b,v2}
      \fmffreeze
      \fmfv{decor.shape=cross,decor.size=2thick,label.dist=4pt,label.angle=90,label=\tiny{$(k_l)$}}{v1}
    \end{fmfgraph*}}
\right)
/
 (\gamma-\alpha_l)_\ext{k} 
\bigg)
+ \sum_{i \notin \gamma \cap \partial \gamma_E}\sum_k \gamma_\ext{k} \otimes \Gamma(i)/\gamma_\ext{k}.
\end{aligned}
\end{equation*}
where in the first line, we have used the fact that $\Gamma
\left( \alpha_m \to 
  \parbox{20pt}{
    \begin{fmfgraph*}(20,20)
      \fmfleft{l}
      \fmfright{r}
      \fmfbottom{b}
      \fmf{plain}{l,v1,v2,v3,r}
      \fmffreeze
      \fmf{doublehead}{b,v2}
      \fmffreeze
      \fmfv{decor.shape=cross, decor.size=2thick,label.dist=4pt,label.angle=90,label=\tiny{$(2)$}}{v3}
    \end{fmfgraph*}}
\right) = \Gamma/\alpha_m (e_m)$, in the above notation.

\bigskip

\noindent Since each vertex 
\begin{fmfgraph*}(20,11)
      \fmfforce{(0w,.15h)}{l}
      \fmfforce{(1w,.15h)}{r}
      \fmf{plain}{l,v,r}
      \fmffreeze
      \fmfv{decor.shape=cross,decor.size=2thick,label.dist=4pt,label.angle=90,label=\tiny{$\ext{0}$}}{v}
\end{fmfgraph*}
adds an electron line to $\Gamma$, we can combine the last three terms to obtain
\begin{equation*}
\begin{aligned}
D(\gamma)&= \sum_\alpha \sum_{m=1}^l \sum_k 
(\alpha_m)_\ext{2} (\gamma-\alpha_m)_\ext{k} \otimes 
\Gamma/ \gamma_\ext{k} (e_m)
+ \sum_k \gamma_\ext{k} \otimes \sum_{\parbox{45pt}{\centering \tiny{$j$ insertion \\points in $\Gamma/\gamma$}}} (\Gamma/\gamma_\ext{k}) (j).
\end{aligned}
\end{equation*}
Thus, we obtain for the sum of $C$ and $D$,
\begin{equation}
\begin{aligned}
\label{cop-i:CD}
C(\gamma) + D(\gamma) &=  \sum_{\alpha,m,k}
\left( \sum_{i \in \alpha_m}(\alpha_m)(i)_\ext{2} +(\alpha_m)_\ext{2} \right)(\gamma-\alpha_m)_\ext{k} \otimes 
\Gamma/ \gamma_\ext{k} (e_m)
\\
&+ \sum_k \gamma_\ext{k} \otimes \sum_j \left( \Gamma/\gamma_\ext{k} \right) (j),
\end{aligned}
\end{equation}
where we recognize the WT-element $W''(\alpha_m)$ for each 1PI self-energy graph $\alpha_m\subset \Gamma$,
\begin{align*}
  W''(\alpha_m)=\sum_{i \in \alpha_m}(\alpha_m)(i)_\ext{1} +(\alpha_m)_\ext{2}.
\end{align*}
Combining the second term in equation \eqref{cop-i:CD} with the two other terms in $\Delta'(W(\Gamma))$ (which involve the coproduct $\Delta'(\Gamma)=\sum_\gamma \gamma_\ext{k} \otimes \Gamma/\gamma_\ext{k}$, it follows that 
$$
\Delta'(W(\Gamma)) = \sum_{\gamma \subset \Gamma}   \sum_{\alpha,m,k} W''(\alpha_m) (\gamma-\alpha_m)_\ext{k} \otimes \Gamma / \gamma_\ext{k} (e_m) + \sum_{\gamma \subset \Gamma} \gamma_\ext{k} \otimes W\left( \Gamma/\gamma_\ext{k} \right),
$$
so that $\Delta'(W(\Gamma))\in I \otimes H + H \otimes I$. 

The other inclusions $\Delta'(W_\mu'(\Gamma)) \in I \otimes H + H \otimes I$ and $\Delta'(W''(\Gamma)) \in I \otimes H + H \otimes I$ follow from the latter equation by the definitions of $W_\mu'(\Gamma)$ and $W''(\Gamma)$ in terms of $W(\Gamma)$. 
\end{proof}

\bigskip

\section{Conclusions}
\label{sect:concl}
We have shown that the Ward-Takahashi identities can be implemented on the Hopf algebra of Feynman graphs of QED as relations that define a Hopf ideal. The quotient Hopf algebra $\tilde H$ by this Hopf ideal is still commutative but has the WT-identities `built in'. The Feynman amplitudes $U$ on $\tilde H$ defined by equation \eqref{eq:feynm-ampl} therefore automatically satisfy the WT-identities in the physical sense ({i.e.} between Feynman amplitudes). Moreover, Theorem \ref{birkhoff} applies and shows that the counterterms and the renormalized Feynman amplitudes are given in terms of the algebra maps $C=U_-: \tilde H \to \cQ$ and $R=U_+ : \tilde H \to \cO$, respectively. In particular, the counterterms as well as the renormalized Feynman amplitudes satisfy the WT-identities. The well-known relation $Z_1=Z_2$ between the renormalization constants as derived by Ward in \cite{War50} is now an easy consequence of their definition in \eqref{eq:ren-const},
\begin{align*}
Z_1 - Z_2 & = \sum_{\Gamma=~
\parbox{20pt}{
\begin{fmfgraph}(20,10)
      \fmfleft{l}
      \fmfright{r}
      \fmf{plain}{l,v}
      \fmf{plain}{v,r}
      \fmfblob{.25w}{v}
\end{fmfgraph}}} 
\sum_{\parbox{45pt}{\centering \tiny{$i$ insertion \\points in $\Gamma$}}} 
C(\Gamma(i)_\ext{1}) + C(\Gamma_\ext{2})\\
&= \sum_{\Gamma} C\left( W''(\Gamma) \right) = 0 .
\end{align*}
In the first line we have used the fact that the contribution $C(\Gamma(i)_\ext{1})$ vanishes whenever $i$ is part of an electron loop, as follows by explicit computation using the Feynman rules (cf. \cite[p.240-241]{PS95}).

\section*{Acknowledgements}
I would like to thank Matilde Marcolli for discussion and remarks, and {\"O}zg{\"u}r Ceyhan for several comments.

\end{fmffile}

\pagebreak
\newcommand{\noopsort}[1]{}


\begin{thebibliography}{10}

\bibitem{BK99}
D.~J. Broadhurst and D.~Kreimer.
\newblock Renormalization automated by {H}opf algebra.
\newblock {\em J. Symbolic Comput.} 27 (1999)  581--600.

\bibitem{BF01}
C.~Brouder and A.~Frabetti.
\newblock Renormalization of {QED} with planar binary trees.
\newblock {\em Eur. Phys. J.} C19 (2001)  715--741.

\bibitem{BF03}
C.~Brouder and A.~Frabetti.
\newblock {QED} {H}opf algebras on planar binary trees.
\newblock {\em J. Algebra} 267 (2003)  298--322.

\bibitem{Col84}
J.~Collins.
\newblock {\em Renormalization}.
\newblock Cambridge University Press, 1984.

\bibitem{CK99}
A.~Connes and D.~Kreimer.
\newblock Renormalization in quantum field theory and the {R}iemann- {H}ilbert
  problem. {I}: {T}he {H}opf algebra structure of graphs and the main theorem.
\newblock {\em Comm. Math. Phys.} 210 (2000)  249--273.

\bibitem{CK00}
A.~Connes and D.~Kreimer.
\newblock Renormalization in quantum field theory and the {R}iemann- {H}ilbert
  problem. {II}: The beta-function, diffeomorphisms and the renormalization
  group.
\newblock {\em Commun. Math. Phys.} 216 (2001)  215--241.

\bibitem{CM04c}
A.~Connes and M.~Marcolli.
\newblock From physics to number theory via noncommutative geometry, part {II}:
  {R}enormalization, the {R}iemann-{H}ilbert correspondence, and motivic
  {G}alois theory.
\newblock hep-th/0411114.

\bibitem{CM04b}
A.~Connes and M.~Marcolli.
\newblock Renormalization and motivic {G}alois theory.
\newblock {\em Int. Math. Res. Not.} 76 (2004) 4073--4091.

\bibitem{CM06}
A.~Connes and M.~Marcolli.
\newblock Quantum fields and motives.
\newblock {\em J. Geom. Phys.} 56 (2006)  55--85.

\bibitem{Dys49}
F.~J. Dyson.
\newblock The {S} matrix in quantum electrodynamics.
\newblock {\em Phys. Rev.} 75 (1949)  1736--1755.

\bibitem{HV73}
\noopsort{HooftVeltman}G. {'t Hooft} and M.~J.~G. Veltman.
\newblock Diagrammar.
\newblock {\em CERN yellow report.} 73 (1973)  1--114.

\bibitem{Kre98}
D.~Kreimer.
\newblock On the {H}opf algebra structure of perturbative quantum field
  theories.
\newblock {\em Adv. Theor. Math. Phys.} 2 (1998)  303--334.

\bibitem{Kre05}
D.~Kreimer.
\newblock Anatomy of a gauge theory.
\newblock {\em Ann. Phys.} Available online 9 March 2006.

\bibitem{KD99}
D.~Kreimer and R.~Delbourgo.
\newblock Using the {H}opf algebra structure of {QFT} in calculations.
\newblock {\em Phys. Rev. D} 60 (1999)  105025, 14.

\bibitem{KY06}
D.~Kreimer and K.~Yeats.
\newblock An \'Etude in non-linear Dyson--Schwinger Equations.
\newblock hep-th/0605096.

\bibitem{PS95}
M.~E. Peskin and D.~V. Schroeder.
\newblock {\em An Introduction to {Q}uantum {F}ield {T}heory}.
\newblock Addison-Wesley, 1995.

\bibitem{VP04}
I.~V. Volovich and D.~V. Prokhorenko.
\newblock Renormalizations in quantum electrodynamics, and {H}opf algebras.
\newblock {\em Tr. Mat. Inst. Steklova} 245 (2004)  288--295.

\bibitem{War50}
J.~C. Ward.
\newblock An identity in quantum electrodynamics.
\newblock {\em Phys. Rev.} 78 (1950)  182.

\end{thebibliography}
\end{document}